\documentclass[12pt]{article}
\pdfoutput=1
\usepackage[english]{babel}
\usepackage[utf8]{inputenc}
\usepackage[T1]{fontenc}
\usepackage{amsmath}
\usepackage{amsthm}
\usepackage{amsfonts}
\usepackage{url}
\usepackage{authblk}
\usepackage{tensor}
\usepackage[bookmarks,colorlinks=false]{hyperref}
\usepackage{graphicx}
\usepackage{mathrsfs}
\usepackage{enumitem}
\usepackage{cite}

\newcounter{mnotecount}[section]

\newcommand{\beq}{\begin{eqnarray}}
\newcommand{\eeq}{\end{eqnarray}}
\newcommand{\ben}{\begin{eqnarray*}}
\newcommand{\een}{\end{eqnarray*}}

\newtheorem*{theorem*}{Theorem}
\theoremstyle{definition}
\newtheorem{defn}{Definition}[section]
\newcommand{\dd}{\mathrm{d}}

\newcommand{\ee}{\operatorname{e}}

\begin{document}
\title{Standing waves in general relativity}
\author[1,3]{Sebastian J. Szybka}
\author[2,3]{Adam Cieślik}
\affil[1]{Astronomical Observatory, Jagiellonian University}
\affil[2]{Institute of Physics, Jagiellonian University}
\affil[3]{Copernicus Center for Interdisciplinary Studies}
\date{}
\maketitle{}
\begin{abstract}
We propose a covariant definition of standing gravitational waves in general relativity.
\end{abstract}

\section{Introduction}

Musical instruments in order to produce a tone use a phenomenon of standing waves. This kind of behavior appears naturally for sound and electromagnetic waves. It may also, although very rarely, appear for water waves. Standing waves are well understood in linear theories (or approximations) where they are obtained as superposition of waves traveling in opposite directions. If nonlinearities are taken into account, the lack of superposition principle complicates studies.

Almost two decades ago Hans Stephani formulated a question \cite{Stephani:2003}: {\it Are there standing gravitational wave solutions of vacuum Einstein's equations?}\footnote{The same year the problem of gravitational standing waves was investigated by Sir Hermann Bondi in his last paper on general relativity \cite{Hermann:2003}. Standing waves appear naturally in some quasistationary approximations in binary black holes inspirals --- an approach introduced by Detweiler \cite{Detweiler} and pursued by Price \cite{Price:2004} and others \cite{Andrade:2003,Beetle:2006,Beetle:2007,Bromley:2005,Hernandez:2009,Mandel:2005}.} In order to answer this question one has first to define what gravitational standing waves are. Unfortunately, their definition cannot be easily generalized from linear theories to nonlinear gravitation. Stephani suggested to look for exact solutions such that \cite{Stephani:2003}
\begin{enumerate}[label=(\Roman*)]
\item The constitutive parts of the metric functions should depend on the timelike coordinate only through a periodic factor, and they should also depend on spacelike coordinates.
\item The time average of some of the metric functions should vanish; in particular, the analogue of the Poynting vector (if there is any) should be divergencefree and the time average of the spatial components should be zero.
\end{enumerate}
The conditions presented above are not covariantly formulated and, as shown by Stephani \cite{Stephani:2003}, not fully satisfactory. 

\section{Standing waves}

In the search for a reliable criterion, we propose to define standing waves using Burnett's \cite{burnett} formulation of the Isaacson high frequency limit \cite{isaacson2} in the form generalized to nonvacuum spacetimes by Green and Wald \cite{greenwald}.

\begin{defn}\label{def}
Let $(M,g)$ be a spacetime satisfying vacuum Einstein's equations. We say that $(M,g)$ contains standing gravitational wave if
\begin{enumerate}[label=(\roman*)]
\item it belongs to a one-parameter family of spacetimes $(M,g(\lambda))$ satisfying the Green-Wald assumptions \cite{greenwald} [we denote the background spacetime with $(M,g^{(0)})$],
\item the Ricci tensor of the background metric $g^{(0)}$ is of a Serge type $[(11)1,1]$ (in Plebański notation $[2S_1-S_2-T]_{(111)}$) with the degenerate eigenvalue equal to zero and remaining eigenvalues $1$, $-1$.
\end{enumerate}
\end{defn}

\noindent
{\it Remark.} 
In a Ricci principal orthonormal tetrad $(x,y,z,\eta)$ of the Serge type $[(11)1,1]$ ($\eta$ is timelike) with the degenerate eigenvalue equal to zero and remaining eigenvalues $1$, $-1$, the effective energy-momentum tensor $t^{(0)}$ (as defined in \cite{greenwald}) may be written as
\begin{equation}\nonumber
t^{(0)}=\rho(x^\alpha)\left(\eta^\flat\otimes\eta^\flat+z^\flat \otimes z^\flat\right)\;,
\end{equation}
where $\rho(x^\alpha)$ is an energy density. An alternative form is
\begin{equation}\nonumber
	t^{(0)}=\frac{1}{2}\rho(x^\alpha)({k^\flat}_+\otimes  {k^\flat}_+ + {k^\flat}_-\otimes {k^\flat}_-)\;,
\end{equation}
where ${k^\flat}_\pm= \eta^\flat\pm z^\flat$. The vectors $k_\pm$ are null (the Green-Wald theorems imply that $t^{(0)}$ is traceless) and oriented in opposite spatial directions. In other words, the effective energy-momentum tensor $t^{(0)}$ is a superposition of null dusts with equal amplitudes, but moving in opposite directions. The background spacetime $(M,g^{(0)})$ may be called effective standing wave spacetime. If the spacetime $(M,g)$ satisfies nonvacuum Einstein equations, then our definition still holds provided that the effective energy-momentum tensor contains contribution from the gravitational radiation $t^{(0)}=t^{(0)}_{GW}+t^{(0)}_{S}$, where $t^{(0)}_{GW}$ satisfies the condition (ii). Such a situation will be illustrated in the example below. Our definition of standing waves may be trivially extended to fields. In the example presented below, standing gravitational waves are accompanied by standing scalar waves.

\section{Example}

The  example\footnote{Stephani \cite{Stephani:2003} refers to this solution as the Kramer solution \cite{Kramer:1999}. Simple redefinitions in this metric lead to totally different physical interpretations of the corresponding spacetime \cite{Kramer:1999}.} 
of the standing gravitational wave presented in the Stephani's paper \cite{Stephani:2003} is a special member of a one-parameter family of exact solutions studied by one of us in the context of the backreaction effect. This one-parameter family corresponds to elementary Einstein-Rosen waves coupled to a massless scalar field. It has been studied within the Green-Wald framework in the article \cite{ers}. Stephani showed \cite{Stephani:2003} that his example satisfies his definition of a `standing wave spacetime'. Our definition of standing waves, if applied to any member of the one-parameter family studied in \cite{ers} (including Stephani's example) leads to the same conclusion in a straightforward manner (both conditions (i), (ii) are satisfied). The effective energy-momentum tensor has the form
\begin{eqnarray}\nonumber
t^{(0)}&=&\frac{\alpha^2+\beta^2}{\pi\rho}\left(\dd t\otimes \dd t+\dd \rho \otimes \dd\rho\right)\\\nonumber
	&=&\frac{1}{2\pi\rho}\left[\alpha^2 ({k^\flat}_+\otimes  {k^\flat}_+ + {k^\flat}_-\otimes {k^\flat}_-) +\beta^2({k^\flat}_+\otimes {k^\flat}_+ + {k^\flat}_-\otimes {k^\flat}_-)\right]\, ,
\end{eqnarray}
where ${k^\flat}_\pm=\dd t\pm \dd\rho$ and $\alpha$, $\beta$ are constant. The parameter $\beta$ controls an amplitude of the scalar field. If $\beta=0$, then the one-parameter family of solutions studied in \cite{ers} satisfies vacuum Einstein's equations. Therefore, the parameter $\alpha$ is related to the amplitude of the gravitational radiation. The effective energy-momentum tensor corresponds to the superposition of ingoing and outgoing gravitational radiation (and to the superposition of ingoing and outgoing waves of the scalar field). This remains true if the solution is rewritten as the three-torus Gowdy cosmology \cite{ers}.

The properties of the background metric confirm our interpretation of solutions studied in \cite{ers}. Although the background metric has a remarkably simple form, it has not been studied extensively in literature before (except the Morgan's article \cite{morgan} where it appears indirectly and the Kramer's article \cite{kramer1} where spherically symmetric equivalent of this metric is derived)
\begin{equation}\label{background}
g^{(0)}=\ee^{\kappa\rho}\left(-\dd t \otimes \dd t+\dd\rho\otimes\dd\rho\right)+\rho^2\dd\varphi\otimes\dd\varphi+\dd z\otimes\dd z\,,
\end{equation}
where $t,z\in(-\infty,+\infty)$, $\rho\in(0,+\infty)$, $\varphi\in[0,2\pi]\;mod\;2\pi$ and $\kappa=2\frac{\alpha^2+\beta^2}{\pi}$ is an auxiliary constant. 

This solution has been discovered in \cite{ers} as a limit of a regular one-parameter family of Einstein-Rosen waves coupled to a massless scalar field. The alternative procedure to derive it follows from generation technique described in \cite{exact} (subsection $25.6.2$). For spacetimes with an orthogonally transitive Abelian group $G_2$ acting on spacelike $2$-surfaces\footnote{In \cite{exact} only $S_2$ is considered.}, pure radiation fields (null dust) may be associated with vacuum solution simply by multiplying the metric coefficient $g_{uv}$ (assuming appropriate parameterization) by the factor $\exp{[F_1(u)+F_2(u)]}$ with arbitrary real functions $F_1(u)$, $F_2(v)$, where $u$ and $v$ are null coordinates. If $F_1=-id$, $F_2=id$, $u=t-\rho$, $v=t+\rho$, ($t$ \nolinebreak is a time coordinate and $\rho$ is a spatial coordinate), then $F_1(u)+F_2(v)=2\rho$. For this particular choice of $F_1$ and $F_2$, a new null dust solution is stationary provided that an original vacuum spacetime was stationary. With the help of this procedure the background metric \eqref{background} may be generated from the Minkowski spacetime in cylindrical coordinates.

The third way to derive the metric \eqref{background} is to superpose outgoing and ingoing null dust Morgan solutions \cite{morgan}. The Morgan solutions are cylindrically symmetric equivalents of Vaidya spacetimes. These solutions, similarly to generalized Einstein-Rosen waves, satisfy superposition principle in the sense that essential part of Einstein equations is linear. The superposition does not work at the level of metric functions, so it is not `complete.' 

In his paper \cite{morgan}, Morgan studied superposed solutions. He stated incorrectly that such spacetime is regular everywhere. The Ricci scalar vanishes as expected, but the Kretschmann scalar blows up at the symmetry axis $\rho=0$
\begin{equation}\nonumber
K=2\left(\frac{2\kappa}{\rho}\right)^2\ee^{-2\kappa\rho}\;.
\end{equation}
This blow up is unexpected because all the metric functions are regular everywhere (only the determinant $det(g^{(0)})$ vanishes at $\rho=0$ which leads to a curvature singularity).

For the one-parameter family of solutions studied in the article \cite{ers}, the C-energy (cylindrical energy as defined by Thorne \cite{thorne}) is constant `on average' for a fixed $\rho$ (at nodes it is strictly constant). Therefore, there is no energy transfer. For the background metric \eqref{background}, the C-energy is equal to $\kappa\rho/4$ and it does not depend on $t$.\footnote{The C-energy corresponds to the time-component of the Poynting vector as has been shown by Stephani \cite{Stephani:2003}.}

In summary, the background metric \eqref{background} corresponds to a cylindrically symmetric spacetime with a central naked singularity accreting and radiating the same amount of gravitational and scalar waves. This spacetime provides an effective description of cylindrical gravitational and scalar standing waves. It is stationary and of Petrov type D. The Weyl scalars are 
\begin{equation}\nonumber
\psi_0=\psi_1=\psi_3=\psi_4=0\;,\quad \psi_2=\frac{1}{12\rho^2}\ee^{-\kappa\rho}\,.
\end{equation}

\section{Stephani's second example}

Stephani illustrates a flaw in his criteria with the following example of the Gowdy universe
\begin{equation}\nonumber
g=\ee^{-2U}\left[ \ee^{2k}(-\dd t\otimes\dd t+\dd\rho\otimes\dd\rho)+\sin^2\rho \sin^2 t \dd\phi\otimes\dd\phi\right]+\ee^{2U} \dd z\otimes\dd z\;,
\end{equation}
where $t\in(0,\pi)$, $\rho,\phi,z\in[0,2\pi]\; mod\;2\pi$ and $U=c \cos\rho\cos t$, $2k=c^2\sin^2\rho\sin^2 t+\ln(\cos^2\rho-\cos^2 t)$ with $c$ being a constant. According to criterion (I), this universe is a standing wave solution. According to criterion (II), it is not. 

The hypersurfaces $\sin(t)=0$ ($t=0$ and $t=\pi$) correspond to initial and final collapse singularities. Singularities at $\sin(\rho)=0$ are only apparent. A natural generalization of this solution to a one-parameter family of metrics is given by the following substitution $t\rightarrow t/\lambda$, $\rho\rightarrow \rho/\lambda$, $c\rightarrow c(\lambda)$, where $c(\lambda)$ is an arbitrary function. Now, for any $c(\lambda)$ the metric is not pointwise convergent to any background metric. Therefore, according to our criteria, there is no reason to believe that this oscillatory solution (contrary to the first example) is a standing wave. This fact support the hypothesis that our definition of standing waves resolved ambiguity in Stephani's criteria.

\section{Summary}

In this paper, we have proposed to define standing gravitational waves in terms of their high frequency limit. We showed that our definition is `nonempty' by providing an example. Moreover, we presented an arguments supporting the claim that ambiguities in Stephani criteria are resolved in our covariant definition. 

It follows from our definition that the effective description of standing gravitational or/and  scalar waves may be provided by superposition of two null dusts of equal density moving in opposite directions. The huge class of solutions to Einstein equations\footnote{The class which was not so well explored so far with notable exceptions in plane \cite{Letelier:1980}, cylindrical \cite{kramer2,Letelier:1994,BookKramer}, and spherical symmetry \cite{kramer1,Date,Gergely:1998,Letelier:1986,PoissonIsrael,BookKramer}.} with the Ricci tensor of the Serge type $[(11)1,1]$ gains a new interesting physical interpretation --- otherwise these solutions would be interpreted mainly as anisotropic perfect fluid solutions which is, of course, much less interesting from the physical standpoint. 	

An interesting open question is: does any effective standing waves spacetime $(M,g^{(0)})$ correspond to some standing wave spacetime $(M,g)$? The results obtained under the assumption of the polarized $U(1)$ symmetry \cite{Huneau:2017led} suggest that it may be the case: any generic local-in-time small-data-polarized-$U(1)$-symmetric solution to the Einstein-multiple null dust system can be achieved as a weak limit of vacuum solutions.
The related question is: does any standing wave spacetime $(M,g)$, as defined by Stephani criterion (II) \cite{Stephani:2003}, belong to a one-parameter family of spacetimes satisfying the Green-Wald assumptions \cite{greenwald}? This question is open and needs further studies.

\vspace{0.3cm}

\noindent{\sc Acknowledgments}

\vspace{0.2cm}

This publication was supported by the John Templeton Foundation
Grant {\it Conceptual Problems in Unification Theories} (No.\ 60671). S.J.S.\ thanks his `musical friends' especially the ensemble {\it The Summer Triangle}. Standing waves created inside their instruments inspired this research.

\bibliographystyle{unsrt}
\bibliography{reportSW}

\end{document}